\documentclass[aps,prl,twocolumn,superscriptaddress]{revtex4}
\usepackage{graphicx}
\usepackage{morefloats}
\usepackage{color}
\usepackage{amssymb}
\usepackage{float}
\usepackage{bm}
\usepackage{amsmath}
\usepackage{natbib}
\usepackage{datetime}

\newdateformat{monthdayyeardate}{%
  \monthname[\THEMONTH]~\THEDAY, \THEYEAR}

\begin{document}

\newcommand{\ham}{\mathcal{H}} 
\newcommand{\Q}{\mathcal{Q}} 

\newcommand{\uvec}[1]{\hat{\mathbf{#1}}}
\newcommand{\plane}[1]{#1}

\title{Planar Hall torque}

\author{C.\,Safranski}
\thanks{These authors contributed equally to this work.}
\author{E. A. Montoya}
\thanks{These authors contributed equally to this work.}
\author{I.\,N.\,Krivorotov}
\email{ilya.krivorotov@uci.edu}
\affiliation{Department of Physics and Astronomy, University of California, Irvine, CA 92697, USA}
\date{\monthdayyeardate\today}


\begin{abstract}
Spin-orbit torques in bilayers of ferromagnetic and nonmagnetic materials hold promise for energy efficient switching of magnetization in nonvolatile magnetic memories. Previously studied spin Hall and Rashba torques originate from spin-orbit interactions within the nonmagnetic material and at the bilayer interface, respectively.  Here we report a spin-orbit torque that arises from planar Hall current in the ferromagnetic material of the bilayer and acts as either positive or negative magnetic damping. This planar Hall torque exhibits unusual biaxial symmetry in the plane defined by the applied electric field and the bilayer normal. The magnitude of the planar Hall torque is similar to that of the giant spin Hall torque and is large enough to excite auto-oscillations of the ferromagnetic layer magnetization.
\end{abstract}

\maketitle

Electric current in bilayers of ferromagnetic (FM) and nonmagnetic (NM) materials can apply torque to the magnetization of the FM \cite{Ando2008}. Such torque arising from spin-orbit coupling (SOC) of conduction electrons can be large enough to drive switching \cite{Miron2011,Liu2012} and auto-oscillations \cite{Liu2012b,Demidov2012,Duan2014,Collet2016,Awad2017} of the FM magnetization. Manipulation of magnetization by spin-orbit torques (SOTs) may be employed in the next generation of magnetic memories and spin torque nano-oscillators (STNOs) that can serve as tunable microwave sources \cite{Slavin2009}. The most studied SOTs in NM/FM bilayers are the spin Hall torque (SHT) \cite{Sinova2015} arising from SOC in the NM layer and the Rashba torque (RT) \cite{Manchon2015}  originating from SOC at the NM/FM interface \cite{Fan2014}. Recently, an unconventional SOT induced by low crystalline symmetry of the NM layer was observed \cite{MacNeill2016a}. Understanding of all types of SOTs that can arise in NM/FM bilayers is needed for formulation of a comprehensive SOT theory and for engineering practical SOT devices.

Here we report an unconventional SOT in NM/FM bilayers that arises from SOC in the FM layer, as opposed to the NM layer or NM/FM interface. This torque modulates magnetic damping of the FM layer in the plane where the effect of SHT and RT on the FM damping is zero. The origin of this SOT can be traced to a spin-polarized electric current in the FM known to give rise to anisotropic magnetoresistance (AMR) and the planar Hall effect (PHE) \cite{Kokado2012}, and thus this SOT can be called the planar Hall torque (PHT). The magnitude of PHT is found to be large enough to cancel magnetic damping of the FM and drive auto-oscillations of the FM magnetization.

\begin{figure*}
\includegraphics[width=\textwidth]{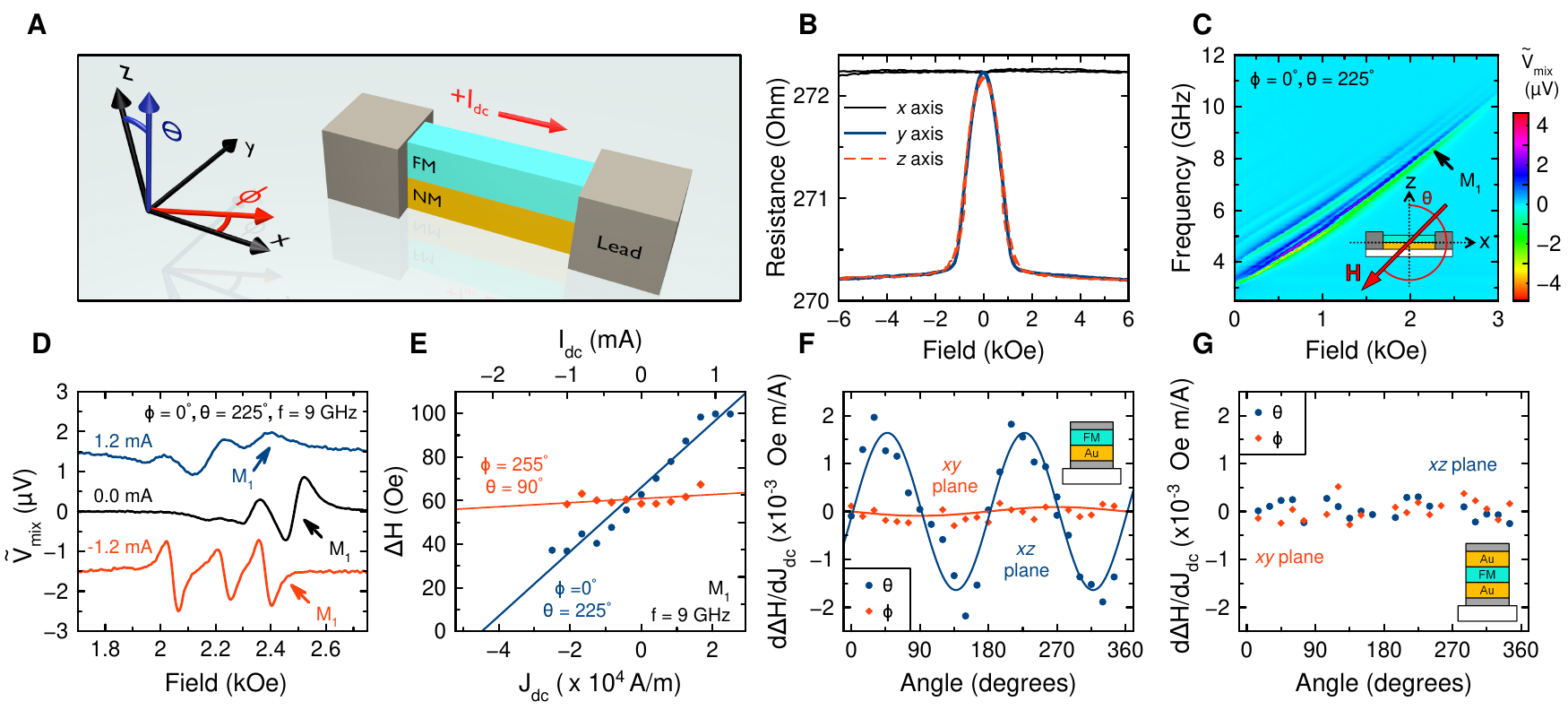}
\centering
\caption{\textbf{ST-FMR measurements at room temperature.} 
(\textbf{A})~Schematic of a NM/FM nanowire device.  (\textbf{B})~Magnetoresistance of the Ta/Au/FM/Ta device. (\textbf{C})~ST-FMR signal as a function of frequency and magnetic field applied in the $xz$ plane at $\theta=225^\circ$. (\textbf{D})~Effect of current bias $I_{\mathrm{dc}}$ on ST-FMR spectra at $\theta=225^\circ$. (\textbf{E})~Linewidth $\Delta H$ of mode $M_1$ as a function of sheet current density $J_{\mathrm{dc}}$ for magnetization in the $xy$ (red) and $xz$ (blue) planes. (\textbf{F},\,\textbf{G})~Angular dependence of $d\Delta H/dJ_{\mathrm{dc}}$ in the $xy$ and $xz$ planes for the Ta/Au/FM/Ta and Ta/Au/FM/Au/Ta devices, respectively.
\label{fig1}}
\end{figure*} 

We measure SOT in NM/FM nanowire devices schematically shown in Fig.\,\ref{fig1}A along with the coordinate system used throughout this report ($\uvec{x}$ is the unit vector along the nanowire, $\uvec{z}$ is the unit vector normal to the bilayer plane). We employ highly resistive Ta seed and cap layers that reduce roughness of the sputter-deposited NM/FM bilayer and prevent its oxidation \cite{methods}. The resulting substrate/Ta(3\,nm)/NM/FM/Ta(4\,nm) multilayers are patterned into  40--50\,nm wide, 40\,$\mu$m long nanowires via e-beam lithography and ion mill etching. Two leads are used to apply electric current over a 80--190\,nm long central part of the nanowire.  We employ a [Co(0.85\,nm)/Ni(1.28\,nm)]$_{2}$/Co(0.85\,nm) superlattice as a FM layer to take advantage of the perpendicular magnetic anisotropy (PMA) in this system \cite{Mangin2006,Arora2017a} that nearly cancels the demagnetizing field of the FM film. The room temperature ($T=295$\,K) AMR curves shown in Fig.\,\ref{fig1}B demonstrate that magnetization of the Ta/Au(3.9\,nm)/FM/Ta nanowire easily saturates for both in-plane and out-of-plane magnetic fields.

We characterize SOTs by field-modulated spin torque ferromagnetic resonance (ST-FMR), which measures the rectified voltage $\tilde{V}_\mathrm{mix}$ generated by the nanowire  at the frequency of the magnetic field modulation \cite{methods,Goncalves2013}. ST-FMR spectra of the Ta/Au(3.9\,nm)/FM/Ta device measured at $T=295$\,K as a function of magnetic field $H$ applied in the $xz$ plane at 45$^\circ$ to the sample normal (Fig.\,\ref{fig1}C) reveal several spin wave resonances in $\tilde{V}_\mathrm{mix} \left( H \right)$. Fig.\,\ref{fig1}D demonstrates that the linewidth of these resonances is altered by a direct current $I_\mathrm{dc}$. We fit the ST-FMR resonances by the magnetic field derivative of the sum of  Lorentzian and anti-Lorentzian functions \cite{Goncalves2013} and plot the linewidth $\Delta H$ of the lowest-frequency mode (labeled $M_1$) as a function of $J_\mathrm{dc}$ in Fig.\,\ref{fig1}E (blue symbols). Here $J_\mathrm{dc}=I_\mathrm{dc}/w$ is the sheet current density, where $w$ is the nanowire width. The linewidth, proportional to the FM magnetic damping $\alpha$, is found to be a linear function of $J_\mathrm{dc}$, which reveals the presence of a SOT that can tune $\alpha$ (referred to as an antidamping SOT in this report). Since the antidamping action of SHT is zero for magnetization lying in the $xz$ plane \cite{Sinova2015}, the data in Fig.\,\ref{fig1}E reveal an unconventional SOT. Furthermore, we find the magnitude of SHT to be small in this system as evidenced by the weak dependence of $\Delta H$ on $J_{\mathrm{dc}}$ for magnetization nearly parallel to the $y$ axis, where the strongest antidamping effect of SHT is expected (red symbols in Fig.\,\ref{fig1}E). 

Fig.\,\ref{fig1}F shows angular dependence of the slope of $\Delta H$ as a function of $J_{\mathrm{dc}}$, which quantifies the strength of an antidamping SOT. For magnetization lying in the $xy$ plane (red symbols), $d\Delta H / d J_{\mathrm{dc}}$ is small and is consistent with a 360$^\circ$ periodicity expected for SHT. In contrast, $d\Delta H / d J_{\mathrm{dc}}$ in the $xz$ plane (blue symbols) is 180$^\circ$-periodic and is large when magnetization makes a 45$^\circ$ angle with respect to the sample normal. The data in Fig.\,\ref{fig1}E show that $\Delta H$ extrapolates to zero at the critical sheet current  density $J_{\mathrm{c}}\approx -4.4 \times 10^{4} $\,A/m, beyond which the FM magnetization is expected to auto-oscillate. 

\begin{figure*}[ptb]
\includegraphics[width=0.75\textwidth]{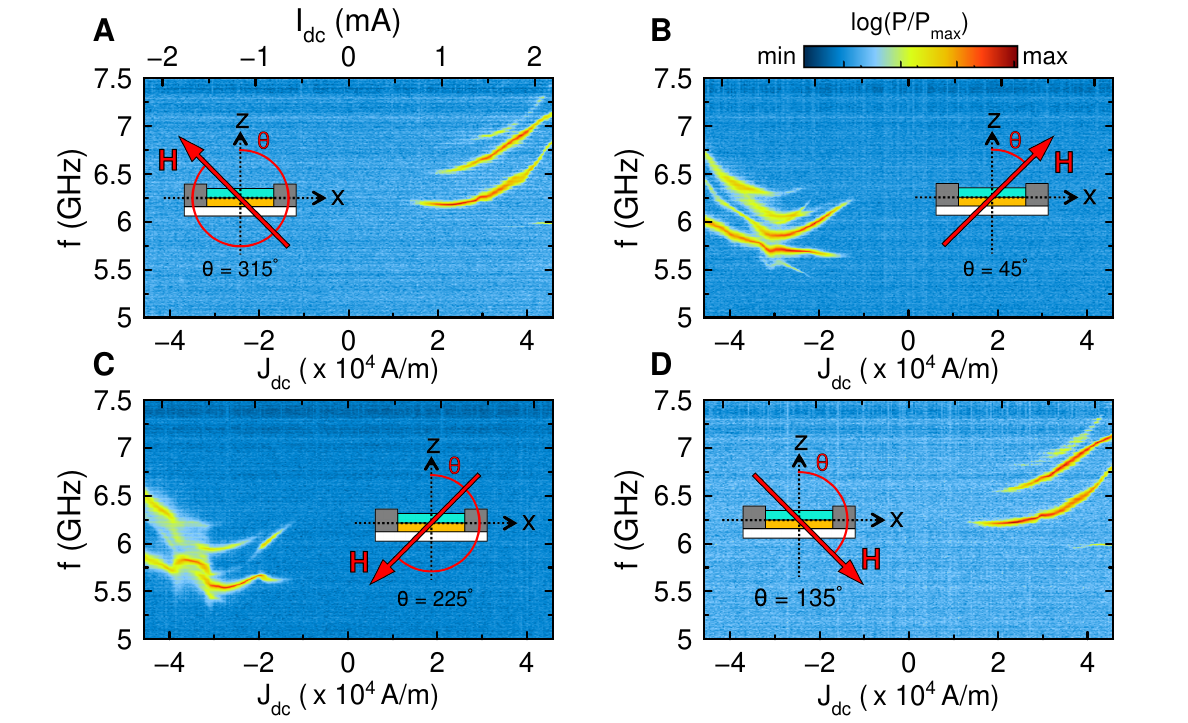}
\centering
\caption{\textbf{ Microwave generation at $\mathbf{T=77}$\,K.} 
Normalized power spectral density of microwave signal generated by the Ta/Au/FM/Ta device in a 1.7\,kOe magnetic field applied in the $xz$ plane at four angles (\textbf{A})~$\theta=315^\circ$, (\textbf{B})~$\theta=45^\circ$, (\textbf{C})~$\theta=135^\circ$, and (\textbf{D})~$\theta=225^\circ$.}
\label{fig2}
\end{figure*}

To reduce ohmic heating, we study the auto-oscillatory dynamics at $T=77$\,K. Fig.\,\ref{fig2} shows spectra of the microwave signal generated by the sample as a function of $J_{\mathrm{dc}}$ for four directions of a 1.7\,kOe magnetic field applied in the $xz$ plane at 45$^\circ$ to the sample normal. The sample generates microwave signal for one current polarity when $|J_{\mathrm{dc}}|$ exceeds $1.7 \times 10^{4}$\,A/m.  The data in Fig.\,\ref{fig2} reveal that $J_\mathrm{c}<0$ for $\theta=45^\circ$ and $\theta=225^\circ$, while $J_\mathrm{c}>0$  for $\theta=135^\circ$ and $\theta=315^\circ$ in agreement with ST-FMR data in Fig.\,\ref{fig1}F. Several auto-oscillatory modes are excited above $J_\mathrm{c}$, with the highest-amplitude mode $M_1$ generating up to 60\,pW of microwave power \cite{methods}.

\begin{figure*}[pt]
\centering
\includegraphics[width=0.75\textwidth]{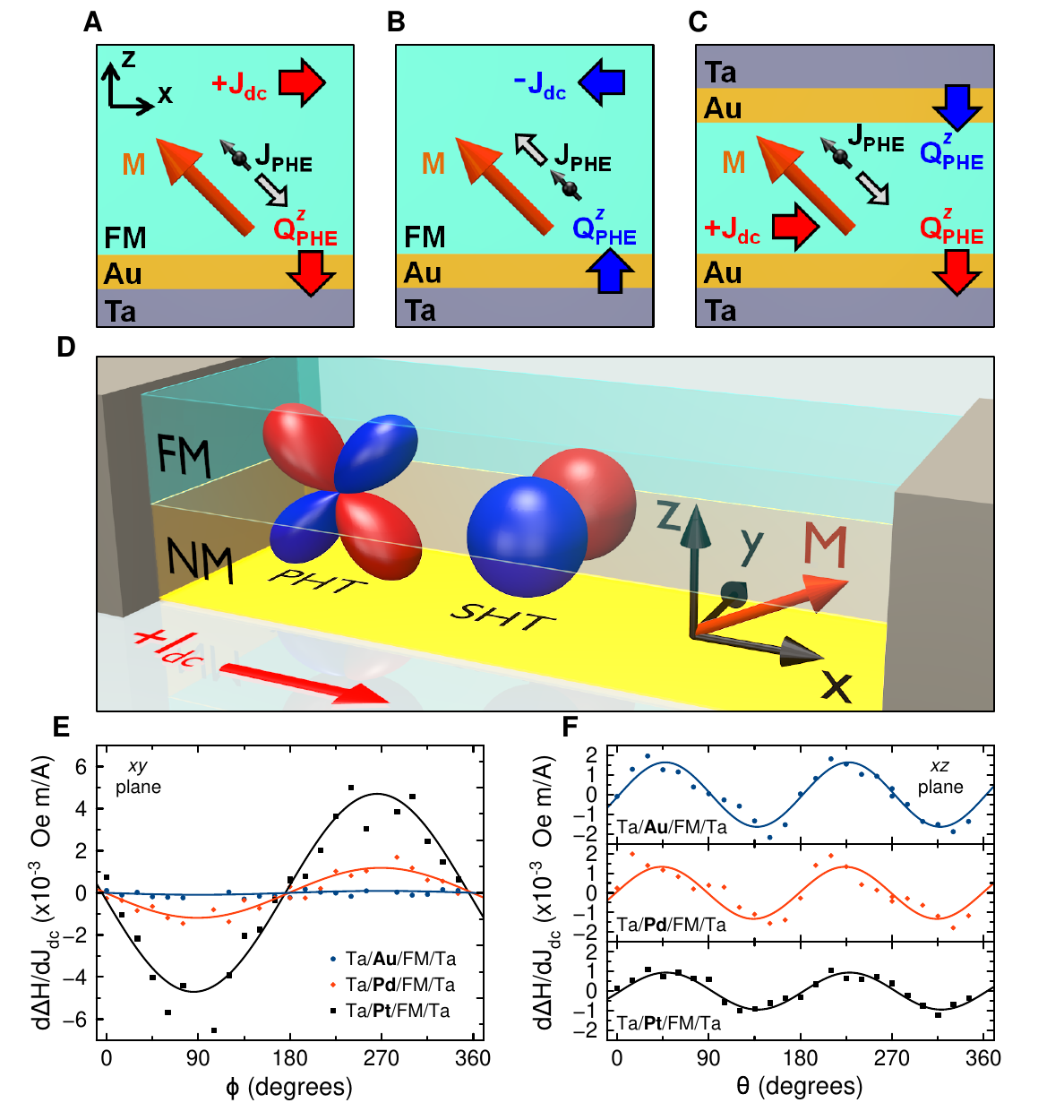}
\caption{\textbf{Planar Hall and spin Hall torques.} 
Schematics illustrating the flow of pure spin current $Q_\mathrm{PHE}^{z}$ driven across the NM/FM interface by spin-polarized planar Hall current $\mathbf{J}_{\mathrm{PHE}}$ in the FM layer for (\textbf{A}) $J_{\mathrm{dc}}>0$ and (\textbf{B}) $J_{\mathrm{dc}}<0$ in the Ta/Au/FM/Ta device and (\textbf{C}) $J_{\mathrm{dc}}>0$ in the Ta/Au/FM/Au/Ta device. (\textbf{D}) Schematic illustrating the angular dependence of antidamping SOTs in NM/FM bilayers: biaxial PHT and uniaxial SHT (red corresponds to negative damping  while blue corresponds to positive damping when $J_{\mathrm{dc}}>0$). $d\Delta H/dJ_\mathrm{dc}$ measured at $T=295$\,K for different NM layers characterizes the strength of antidamping SHT in the $xy$ plane (\textbf{E}) and antidamping PHT in the $xz$ plane (\textbf{F}).}
\label{fig3}
\end{figure*}

 The angular dependence of $d\Delta H / d J_{\mathrm{dc}}$ in Fig.\,\ref{fig1}F is well fit by $\cos(\theta)\sin(\theta)\cos(\phi)=(\uvec{m} \cdot \uvec{x}) (\uvec{m} \cdot \uvec{z})$ (blue curve), where $\uvec{m}$ is the unit vector in the direction of the FM magnetization. We argue that this biaxial symmetry of the antidamping SOT can arise from the planar Hall current generated by SOC in FM conductors \cite{Kokado2012,Taniguchi2015}.  The planar Hall effect results in a current of spin-polarized electrons of density $\mathbf{J}_{\mathrm{PHE}}=\Delta \sigma_{\mathrm{AMR}}(\uvec{m} \cdot \mathbf{E}) \uvec{m}$ flowing parallel to the FM magnetization, where $\mathbf{E}$ is the electric field applied to the FM ($\mathbf{E}\approx E\uvec{x}$) and $\Delta\sigma_{\mathrm{AMR}}$ is anisotropic part of the FM conductivity. Assuming the antidamping SOT originates from exchange of angular momentum between the FM and NM layers via spin currents, only the component of $\bf J_{\mathrm{PHE}}$ normal to the NM/FM interface can contribute to the SOT: 
\begin{equation} 
\label{eq:Jz}
J_{\mathrm{PHE}}^z=\Delta \sigma_{\mathrm{AMR}}E(\uvec{m} \cdot \uvec{x}) (\uvec{m} \cdot \uvec{z}).
\end{equation}
The angular dependence of $J_{\mathrm{PHE}}^z$ is identical to the measured angular dependence of the antidamping SOT shown in Fig.\,\ref{fig1}F, which suggests that this SOT originates from the planar Hall current. When $J_{\mathrm{PHE}}^z<0$, the planar Hall current drives electrons with magnetic moments aligned with the FM magnetization (black arrows) from the FM layer into the NM layer \cite{Kokado2012}, as shown in Fig.\,\ref{fig3}A. In a steady state, the net electron current across the NM/FM interface is zero, which implies a backflow of electrons from the NM layer to the FM layer. The backflow current from a NM layer with strong spin flip scattering is weakly spin polarized, which results in a net pure spin current density $Q_\mathrm{PHE}^{z}$ flowing in the $-\uvec{z}$ direction across the NM/FM interface which acts as negative damping. 
Such a spin current changes between negative and positive damping under change in the sign of $J_\mathrm{dc}$ as shown Fig.\,\ref{fig3}B, where $Q_\mathrm{PHE}^{z}$ now flows in the $+\uvec{z}$ direction across the NM/FM interface. Since $Q_\mathrm{PHE}^{z}$ changes sign upon a $90^\circ$ rotation of $\uvec{m}$ in the $xz$ plane (Eq.~(\ref{eq:Jz})), the effect of PHT changes from positive to negative damping upon such a rotation in agreement with the data in Fig.\,\ref{fig1}F and Fig.\,\ref{fig2}.

Fig.\,\ref{fig3}D shows angular dependence of the SOT-induced damping for PHT $(\uvec{m} \cdot \uvec{x}) (\uvec{m} \cdot \uvec{z})$ and SHT $(\uvec{m} \cdot \uvec{y})$, both allowed by symmetry \cite{Garello2013}. Red and blue in Fig.\,\ref{fig3}D represent negative damping and positive damping torques for $J_\mathrm{dc}>0$, respectively.  The biaxial angular symmetry of PHT is clearly illustrated by this figure. We also expect PHT to be zero in symmetric NM/FM/NM trilayers because the net spin current flowing across the two NM/FM interfaces is zero as illustrated in Fig.\,\ref{fig3}C. Indeed, our measurements of antidamping SOT in a Ta/Au/FM/Au/Ta trilayer nanowire displayed in Fig.\,\ref{fig1}G demonstrate that PHT in this system is negligibly small.

To understand the effect of the NM layer on antidamping SOTs, we measure PHT and SHT in a series of Ta/NM/FM/Ta nanowires that employ different NM layers with similar sheet resistances: Au(3.9\,nm), Pt(7.0\,nm), Pd(8.0\,nm) \cite{methods}. The angular dependence of $d\Delta H / d J_{\mathrm{dc}}$ quantifying an antidamping SOT is measured in the $xy$ and $xz$ planes. Fig.\,\ref{fig3}E shows that the magnitude of the $360^\circ$-periodic antidamping SHT in the $xy$ plane strongly depends on the NM material, in agreement with previous studies \cite{Mosendz2010}. The data in Fig.\,\ref{fig3}F demonstrates that the magnitude of the $180^\circ$-periodic PHT is nearly the same for all three NM layers, consistent with PHT arising from spin current generated by the FM layer rather than the NM layer, which merely serves as a good spin sink \cite{methods}. Fig.\,\ref{fig3}E and  Fig.\,\ref{fig3}F clearly demonstrate that the magnitudes of SHT and PHT are not correlated, which further supports the difference in the mechanisms giving rise to these SOTs. 

It is expected from Eq.~(\ref{eq:Jz}) that the magnitude of PHT is proportional to the magnitude of AMR in the FM layer. We find that the AMR ratio in our Ta/Au/FM/Ta multilayer increases by a factor of 1.8 upon cooling the sample from 295 K to 77 K \cite{methods}.  At the same time, the critical sheet current density $J_{\rm{c}}$ decreases by a factor of 1.8 (from  $-4.4 \times 10^{4}$\,A/m at 295\,K to $-2.5 \times 10^{4}$\,A/m at 77\,K as measured by ST-FMR at 9\,GHz \cite{methods}). As $J_{\rm{c}}$ is inversely proportional to the PHT magnitude, these data strongly support the proportionality of PHT and AMR. The strong increase of the PHT magnitude upon cooling is in sharp contrast to that of SHT known to decrease with decreasing temperature \cite{Duan2014}.

Recently, PHT and the anomalous Hall torque (AHT) were predicted in FM/NM/FM trilayers, where the planar and anomalous Hall currents generated in one FM layer apply torques to magnetization of the other FM layer \cite{Taniguchi2015}. Experiments show that current-driven coupling between two FM layers in a FM/NM/FM trilayer can be achieved via SOTs generated at the NM/FM interfaces \cite{Humphries2017}. Our work demonstrates that a strong antidamping SOT originating from current in the FM layer and acting on magnetization of the same FM layer can arise in a NM/FM bilayer. 

While our data demonstrate that the PHT symmetry and magnitude are determined by the symmetry and magnitude of the planar Hall current in the FM, the microscopic mechanism giving rise to PHT requires further theoretical understanding. Spin polarization of the planar Hall current $\uvec{p}$ in the FM is collinear with the FM magnetization, and thus cannot give rise to the conventional SOTs proportional to $\uvec{m} \times \uvec{p}$ or $\uvec{m} \times (\uvec{p} \times \uvec{m})$ \cite{Garello2013}. It is, however, possible that $\uvec{p}$ rotates upon traversing the NM/FM interface and develops a component transverse to $\uvec{m}$, which can generate an antidamping SOT. Such rotation of the current polarization can arise from interfacial SOC scattering \cite{Amin2016b} or spin swapping \cite{Saidaoui2016}. Alternatively, antidamping PHT can be produced from a longitudinal spin current ($\uvec{p}\parallel\uvec{m}$) via a magnonic mechanism similar to that discussed and demonstrated in the context of spin Seebeck torque in NM/FM bilayers \cite{Bender2016,Safranski2016b}. Further theoretical and experimental work is needed to understand the microscopic mechanism of PHT.

In summary, we observe a biaxial antidamping spin-orbit torque arising from planar Hall current in bilayers of ferromagnetic and nonmagnetic metals, and demonstrate operation of a spin torque nano-oscillator driven by this planar Hall torque. We expect the planar Hall torque to play a significant role in spin-orbit torque switching of magnetization \cite{Mann2017} as well as in current-driven domain wall \cite{Ryu2013,Duttagupta2017a} and skyrmion \cite{Legrand2017} motion in bilayers of ferromagnetic and nonmagnetic materials.

We thank M. Arora and E. Girt for helpful discussion on Co/Ni multilayer growth. Work of E.A.M. on optimization and deposition of magnetic multilayers was supported as part of the Spins and Heat in Nanoscale Electronic Systems (SHINES), an Energy Frontier Research Center funded by the US Department of Energy, Office of Basic Energy Sciences (BES) under Award No. DE-SC0012670. Nanowire device fabrication done by C.S. and E.A.M. was supported by the U.S. Department of Energy, Office of Basic Energy Sciences under Award No. DE-SC0014467. Spin torque oscillator measurements performed by C.S. and E.A.M were supported by the National Science Foundation under Award No. DMR-1610146. Work of C.S. on rotatable-field ST-FMR setup development and fabrication was supported by the National Science Foundation under Award No. EFMA-1641989. ST-FMR measurements made by  C.S.  and E.A.M. were supported by the Army Research Office under Award No. W911NF-16-1-0472. Work of E.A.M. on absorptive FMR and spin pumping measurements was supported by the Defense Threat Reduction Agency under Award No. HDTRA1-16-1-0025. Work of I.N.K. on experiment design and SOT analysis was supported by the National Science Foundation under Award No. ECCS-1708885. All authors analyzed the data and co-wrote the paper.  All data described in this report are available from the corresponding author on request. The authors declare no competing financial interests.





\clearpage
\widetext

\begin{center}
\textbf{\large Supplemental Materials: Planar Hall torque}
\bigskip
\end{center}
\onecolumngrid
\setcounter{equation}{0}
\setcounter{figure}{0}
\setcounter{table}{0}
\setcounter{page}{1}
\renewcommand{\theequation}{S\arabic{equation}}
\renewcommand{\thefigure}{S\arabic{figure}}
\renewcommand{\thetable}{S\arabic{table}}

\renewcommand{\figurename}{Supplementary Figure}
\renewcommand\refname{Supplementary References}
\renewcommand\tablename{Supplementary Table}
\def\bibsection{\section*{\refname}}

\section{Multilayer deposition}

The multilayers are  deposited by  magnetron sputtering on Al$_2$O$_3\left( 0001 \right)$ substrates in 2 mTorr Ar with a base pressure $\leq 3.0 \times 10^{-8}$ Torr. The film stacks consist of ferromagnetic metal (FM) and nonmagnetic metal (NM) layers and generally have either of the following structure types:
\begin{enumerate}
\item{Al$_2$O$_3\left( 0001 \right)$/Ta(3 nm)/NM/FM/Ta(4 nm)} \label{NMFM}
\item{Al$_2$O$_3\left( 0001 \right)$/Ta(3 nm)/NM/FM/NM/Ta(4 nm)} \label{NMFMNM}
\end{enumerate}
Since the Ta(3 nm) seed layer and Ta(4 nm) capping layer are common to all structures, we refer to type 1 as NM/FM and type 2 as NM/FM/NM structures. The Ta seed layer is employed to promote growth of a smooth multilayer \cite{Arora2017} and the Ta capping layer prevents the multilayer oxidation. The composite FM layer is a superlattice of exchange coupled Co and Ni layers: Co(0.85 nm)/Ni(1.28 nm)/Co(0.85 nm)/Ni(1.28 nm)/Co(0.85 nm). We have chosen the Co/Ni superlattice due to its large PHE and AMR \cite{Mcguire1975} as well as its significant perpendicular magnetic anisotropy (PMA) \cite{Daalderop1992,Arora2017a}. The thicknesses of the Co and Ni layers in the composite FM are chosen to nearly balance the easy plane magnetic shape anisotropy of the FM film by PMA in order to be able to saturate the FM magnetization in any direction by a small magnetic field. We employ Pt, Pd, and Au as the NM layers. In type 1 multilayers, we use NM = Pt(7.0\,nm), Pd(8.0\,nm), or Au(3.9\,nm), with the NM thickness chosen to keep the NM layer sheet resistance nearly the same. For type 2 multilayer, we employ Au(2.5\,nm) as the bottom NM material and Au(1.5\,nm) as the top NM layer.  The thickness of the top and bottom NM layers are chosen to be different in order to generate a non-zero microwave Oersted field applied to the FM magnetization in ST-FMR measurements, which strongly enhances the amplitude of the magnetic oscillations and the signal-to-noise ratio of the ST-FMR signal.

\section{Sheet resistance measurements}

\begin{figure}[!htb]
	\begin{center}
		\includegraphics[width=0.5\textwidth]{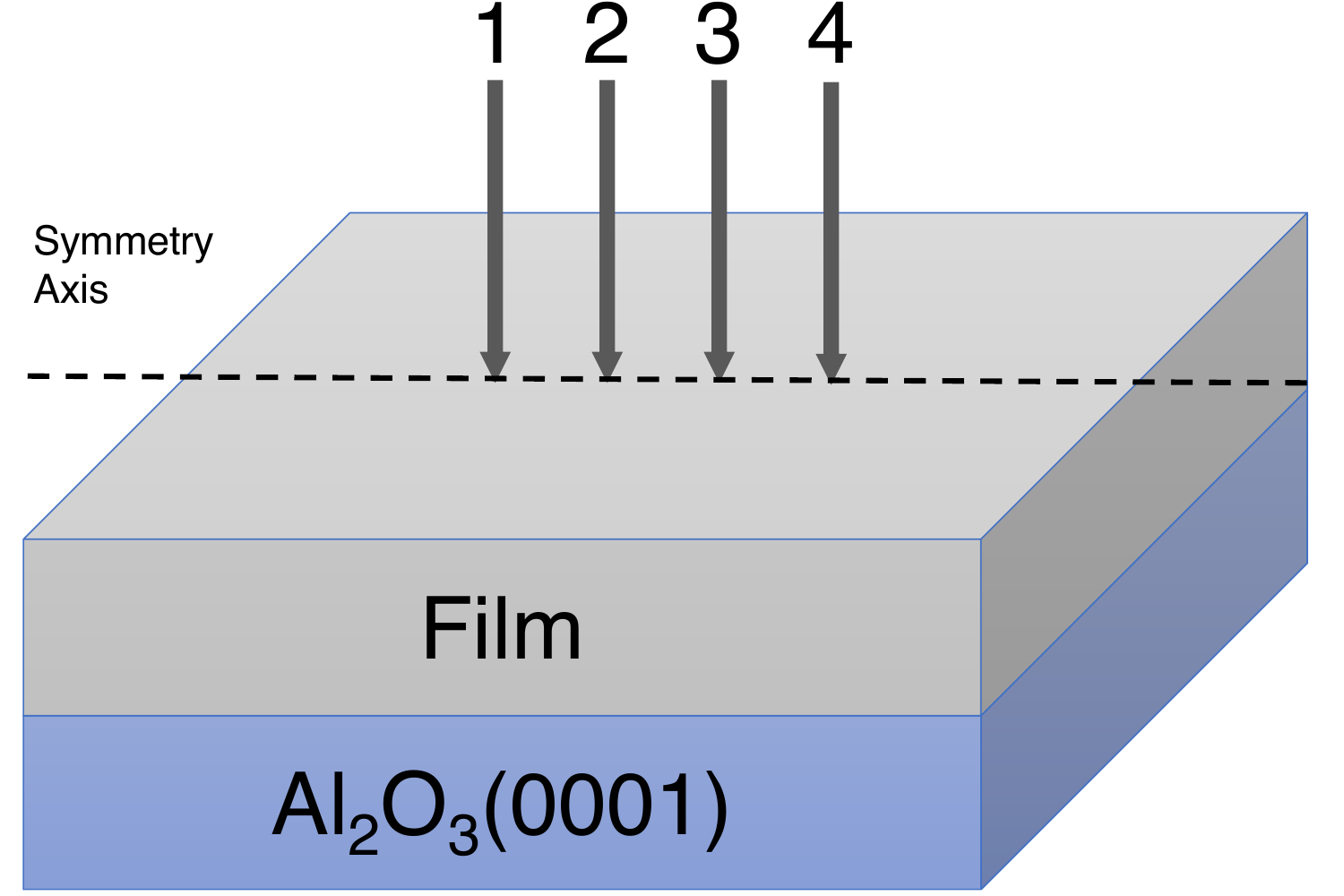}
	\end{center}
\caption{
Schematic of 4-point probe measurement.
} \label{fig:4PointProbe}
\end{figure}

We make sheet resistance measurements of magnetic multilayers by the collinear four point probe technique \cite{Miccoli2015}. Four equidistant probes make contact along the symmetry axis to $ 10 \times 10 $ mm$^2$ film samples, as shown in Fig. \ref{fig:4PointProbe}. The resistances $ R_{A} =V_{23}/I_{14} $, $ R_{B} =V_{24}/I_{13} $, and $ R_{C} =V_{43}/I_{12} $ are measured by means of 4 terminal source-meter. The subscripts on $I$ and $V$ indicate the probes where current is supplied and voltage is measured; for example, for  $ R_{A}$, a direct current $I$ is supplied between points 1 and 4 and the voltage is measured at points 2 and 3. $ R_{A}$, $R_{B} $, and $ R_{C}$ must satisfy the following two relations \cite{Thorsteinsson2009}:
\begin{equation}\label{eq:4Point1}
\exp \left( -{\frac{2 \pi R_A }{R_{\mathrm{S}}}} \right) + \exp \left( -{\frac{2 \pi R_C }{R_{\mathrm{S}}}} \right) =1
\end{equation}	
and
\begin{equation}\label{eq:4Point2}
\exp \left( {\frac{2 \pi R_A }{R_{\mathrm{S}}}} \right) - \exp \left( {\frac{2 \pi R_B }{R_{\mathrm{S}}}} \right) =1,
\end{equation}
where $R_{\mathrm{S}}$ is the sheet resistance of the film.
Small deviations in probe spacing are corrected by averaging a symmetric measurement for the $R$s, and thermal voltages/offsets are accounted for by switching voltage leads, therefore: 
\begin{equation}
	\begin{array}{ll}
		R_{A} &=  \left( V_{23}/I_{14} + |V_{32}|/I_{14} \right)  / 2 \\
		R_{B} &=  \left( V_{24}/I_{13} + V_{13}/I_{24} + |V_{42}|/I_{13} + |V_{31}|/I_{24} \right)  / 4 \\
		R_{C} &=  \left( V_{43}/I_{12} + V_{12}/I_{43} + |V_{34}|/I_{12} + |V_{21}|/I_{43} \right)  / 4.
	\end{array}
\end{equation}

The measured sheet resistances of various multilayer films are summarized in Table \ref{tbl:SheetResistanceSummary}. The current densities in each layer can be determined from the  parallel resistor model. The top Ta(4\,nm) is taken to have the same resistivity as the bottom Ta(3\,nm) because approximately 1-2 nm of the top Ta layer oxidizes under ambient conditions \cite{Montoya2016}. The data in Table \ref{tbl:SheetResistanceSummary} demonstrate that the Ta layers have much higher sheet resistances than other layers in the multilayer stack and thus the electric current density in the Ta layers is small in all our measurements.

\begin{table*}[h]
	\centering
  	\caption{Summary of film level sheet resistances determined by 4 point probe measurements. Numbers in parenthesis are layer thickness in nm.}\label{tbl:SheetResistanceSummary}
  	\begin{tabular}{lc}
	\hline
 		Sample 																& $R_{\mathrm{S}} \left( \Omega \right)$ \\
        \hline
 		Ta(3.0) 														 	& 929 			\\
        Ta(3.0)/Au(3.9) 												 	& 27.0 \\
 		Ta(3.0)/Pt(7.0)													 	& 28.2 \\
 		Ta(3.0)/Pd(8.0)													 	& 27.9 \\
 		Ta(3.0)/Au(3.9)/[Co(0.85)/Ni(1.28)]$_2$/Co(0.85)/Ta(4)				& 19.3 \\
 		Ta(3.0)/Pt(7.0)/[Co(0.85)/Ni(1.28)]$_2$/Co(0.85)/Ta(4)				& 20.5 \\
 		Ta(3.0)/Pd(8.0)/[Co(0.85)/Ni(1.28)]$_2$/Co(0.85)/Ta(4)				& 17.3 \\
   		Ta(3.0)/Au(2.5)/[Co(0.85)/Ni(1.28)]$_2$/Co(0.85)/Au(1.5)/Ta(4)		& 22.2 \\
        \hline
  \end{tabular}
\end{table*}

\section{Broadband ferromagnetic resonance}

We employ a conventional broadband ferromagnetic resonance technique (FMR) to measure magnetic damping in the NM/FM and NM/FM/NM multilayer films in order to characterize spin pumping across the NM/FM interfaces and spin sink efficiencies of the NM layers. FMR measurements are carried out at room temperature using a broadband microwave generator,  coplanar waveguide, and planar-doped detector diode. The measurements are performed at discrete frequencies in an in-plane field-swept, field-modulated configuration, as detailed in Ref.~\cite{Montoya2014}. The measured FMR data are described by an admixture of the $\chi^{'}$ and $\chi^{''}$ components of the  complex transverse magnetic susceptibility, $ \chi = \chi^{'} + i \chi^{''} $. The FMR data are fit as described by Ref.~\cite{Montoya2014, Montoya2016a}.

The Landau-Lifshitz-Gilbert (LLG) equation describes the magnetization dynamics in the FMR experiment:
\begin{equation}\label{eq:LLG}
 \frac{ \partial \mathbf{M} }{ \partial t } = -\gamma \left[ \mathbf{M} \times \mathbf{H}_{\mathrm{eff}}  \right]  +  \alpha \left[ \mathbf{M} \times \frac{ \partial \uvec{m} }{ \partial t }  \right],
\end{equation}	
where $ \mathbf{M} $ is the instantaneous magnetization vector with magnitude  $ M_{\rm{s}} $, $ \uvec{m} $ is the unit vector parallel to $ \mathbf{M} $, $\mathbf{H}_{\rm{eff}}$ is the sum of internal and external magnetic fields, $\gamma = g \mu_{\mathrm{B}}/\hbar $ is the absolute value of the gyromagnetic ratio, $g$ is the Land\'{e} g-factor, $\mu_{\mathrm{B}}$ is the Bohr magneton, $\hbar$ is the reduced Planck constant, and $\alpha$ is the dimensionless Gilbert damping parameter.

\begin{figure*}[!htb]
\center
\includegraphics[width=.9\textwidth]{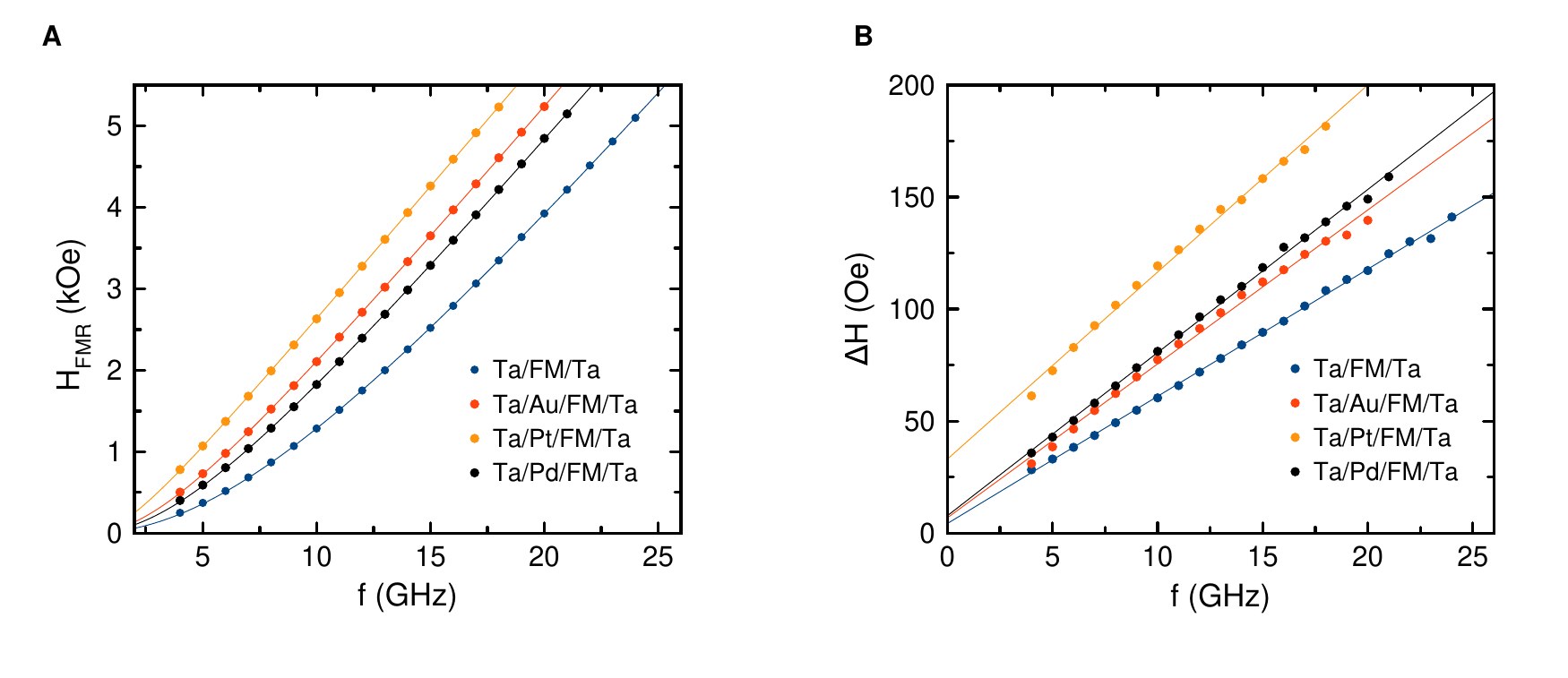}
\caption{
(\textbf{A}) Ferromagnetic resonance field as a function of frequency for films measured in the in-plane magnetic field configuration. (\textbf{B}) FMR linewidth $\Delta H$ as a function of frequency.}
\label{fig:FMR_film}
\end{figure*}

The samples studied are polycrystalline and show negligible in-plane anisotropy. The in-plane resonance condition is
\begin{equation}\label{eq:Resonance}
\left( \frac{\omega}{\gamma} \right)^2 = \left( H_{\mathrm{FMR}} \right)\left( H_{\mathrm{FMR}} + 4 \pi M_{\mathrm{eff}} \right),
\end{equation}	
where $\omega =2 \pi f $ is the microwave angular frequency, $H_{\mathrm{FMR}}$ is the resonance field, $ 4\pi M_{\mathrm{eff}} = 4 \pi M_{\mathrm{s}} - 2 K_{\mathrm{U}} / M_{\mathrm{s}} $, and $ K_{\mathrm{U}} $ is the perpendicular-to-film-plane uniaxial anisotropy. Figure~ \ref{fig:FMR_film}A shows example data of $H_{\mathrm{FMR}}$ as a function of frequency. The data are fit using equation~\ref{eq:Resonance} to extract  $4 \pi M_{\mathrm{eff}}$ and $g$, which are tabulated in Table~\ref{tbl:FilmFMRSummary}.

The measured FMR linewidth defined as half-width of the resonance curve is well described by Gilbert-like damping,
\begin{equation}\label{eq:GilbertDamping}
\Delta H(\omega)= \alpha \frac{\omega}{\gamma} + \Delta H(0),
\end{equation}	
where  $ \omega $ is the microwave angular frequency, and  $ \Delta H(0) $ is the zero-frequency line broadening due to long range magnetic inhomogeneity \cite{Heinrich1993,McMichael2003,Bland2005}. Figure \ref{fig:FMR_film}B shows example data of $\Delta H$ as a function of frequency. The data are fit using equation~\ref{eq:GilbertDamping} to extract the parameters $\alpha$ and $ \Delta H(0)$, which are tabulated in Table~\ref{tbl:FilmFMRSummary}.

\begin{table*}
	\centering
  	\caption{Summary of film level magnetic properties determined by broadband ferromagnetic resonance.  Numbers in parenthesis are layer thickness in nm. The composite ferromagnet is FM=[Co(0.85)/Ni(1.28)]$_2$/Co(0.85).}\label{tbl:FilmFMRSummary}
  	\begin{tabular}{lcccc}
	\hline
 		Sample 		   				& $4\pi M_{\mathrm{eff}} $ (Oe) & $g$  & $\alpha$ ($10^{-3}$)	& $\Delta H(0)$ (Oe)  \\
        \hline
 		Ta(3.0)/FM/Ta(3.0)	   		& 7160                          & 2.17 & 17.2 				& 4 		\\
 		Ta(3.0)/Au(3.9)/FM/Ta(3.0)    	& 3010    				       & 2.17 & 20.9 				& 7  		\\
 		Ta(3.0)/Au(2.5)/FM/Au(1.5)/Ta(3.0) 	& 3220    				       & 2.14 & 21.7 				& 4 		\\
 		Ta(3.0)/Pt(7.0)/FM/Ta(3.0)    	& 1480    				       & 2.17 & 25.3 				& 33 		\\
 		Ta(3.0)/Pd(8.0)/FM/Ta(3.0)    	& 4060    				       & 2.18 & 22.2   				& 8 		\\
        \hline
  \end{tabular}
\end{table*}

The values of $\alpha$ in Table~\ref{tbl:FilmFMRSummary} show that the Ta/FM/Ta sample has the lowest damping. This means that the spin pumping is smallest at the Ta/FM and FM/Ta interfaces \cite{Urban2001,Simanek2003,Tserkovnyak2005}. These data imply that spin current density across the FM/Ta interface is low and thus PHT arising from net spin current across this interface is small. Adding a Au insertion layer at the bottom Ta/FM interface, Ta/FM/Ta $\rightarrow$ Ta/Au/FM/Ta, shows a significant increase in the spin pumping induced damping. It is noteworthy that the thickness of Au(3.9 nm) is much less than the spin diffusion in Au (30-60 nm) suggesting that the Au/Ta bilayer acts as an efficient spin sink. Furthermore, adding another Au layer between the top FM/Ta interface shows further increase in spin pumping induced damping. This shows that the Au layer in the Au/Ta bilayer facilitates efficient spin transfer from the FM layer. The values of $\alpha$ in the Pt/FM and Pd/FM structures are high indicating that the Pt and Pd layers are efficient spin current sinks.

The result of low damping at the FM/Ta interface is somewhat surprising as much literature has shown that many FM/Ta exhibit significant spin pumping induced damping \cite{Hahn2013,Montoya2016}, i.e. Ta acts as an efficient spin sink when adjacent to a FM. However, there are other cases where spin pumping directly into Ta did not occur. In one of the earliest spin pumping experiments on NM/NiFe/NM films,  Mizukami et. al. \cite{Mizukami2001} showed large interface damping when NM = Pt, Pd, but no interface damping when NM = Ta. A similar effect was seen by Liu et. al.~\cite{Liu2012} at the Ta/CoFeB interface. More recently, Singh et. al. \cite{Singh2017a}, have shown spin pumping experiments in Ta/NM1/Co/NM2/Ta, Ta/Co/Ta, and Cu/Ta/Cu samples, where NM1 and NM2 are various combinations of Pt and Au layers. Their results are consistent with our data in Table \ref{tbl:FilmFMRSummary} and show, that under certain sample growth conditions, spin transfer from Co to Ta at the Co/Ta interfaces is inefficient.

\section{Spin torque ferromagnetic resonance}
\begin{figure*}[!htb]
\centering
\includegraphics[width=0.5\textwidth]{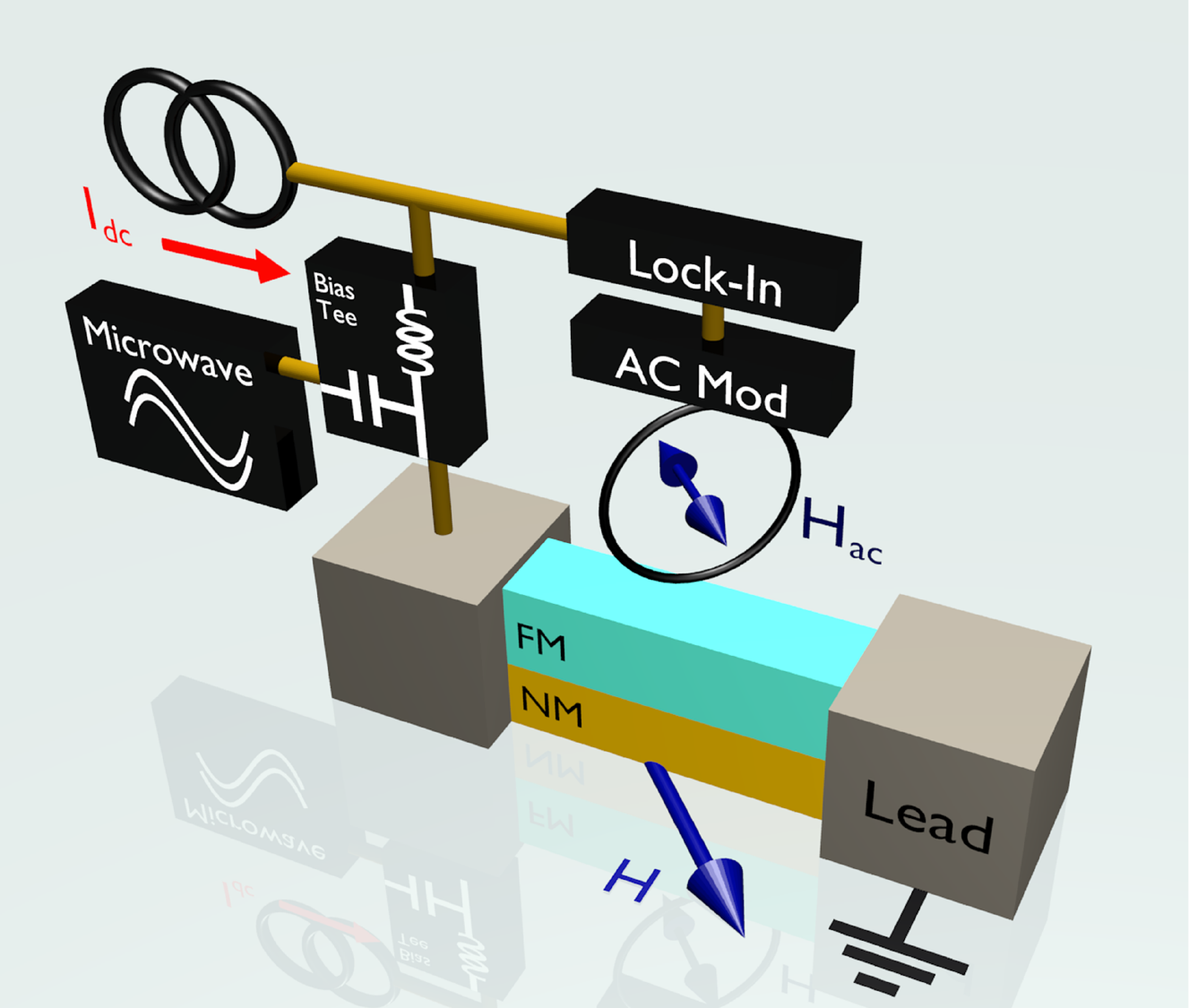}
\caption{Schematic of the setup for ST-FMR measurements with magnetic field modulation.}
\label{fig:STFMR}
\end{figure*}

We measure the ferromagnetic resonance linewidth as a function of direct current bias in the nanowire devices by spin-torque ferromagnetic resonance (ST-FMR) \cite{Tulapurkar2005, Sankey2006}. In this method, a rectified voltage $V_{\mathrm{mix}}$ generated by the sample in response to the applied microwave current is measured as a function of the drive frequency and external magnetic field \cite{Chiba2015}. Resonances in $V_{\mathrm{mix}}$ are observed at the frequency and field values corresponding to spin wave eigenmodes of the system \cite{Ando2009}. To improve the sensitivity of the method, we modulate the applied magnetic field and use lock-in detection to measure ac voltage $\tilde{V}_\mathrm{mix}$ generated by the sample at the frequency of the magnetic field modulation as illustrated in Fig.\,\ref{fig:STFMR} \cite{Goncalves2013}. This signal is proportional to magnetic field derivative of the rectified voltage $ \tilde{V}_\mathrm{mix}\left(H\right) \sim dV_{\mathrm{mix}}\left(H\right) / dH $.

\section{Generated microwave power and the critical current}

\begin{figure*}[!h]
\includegraphics[width=\textwidth]{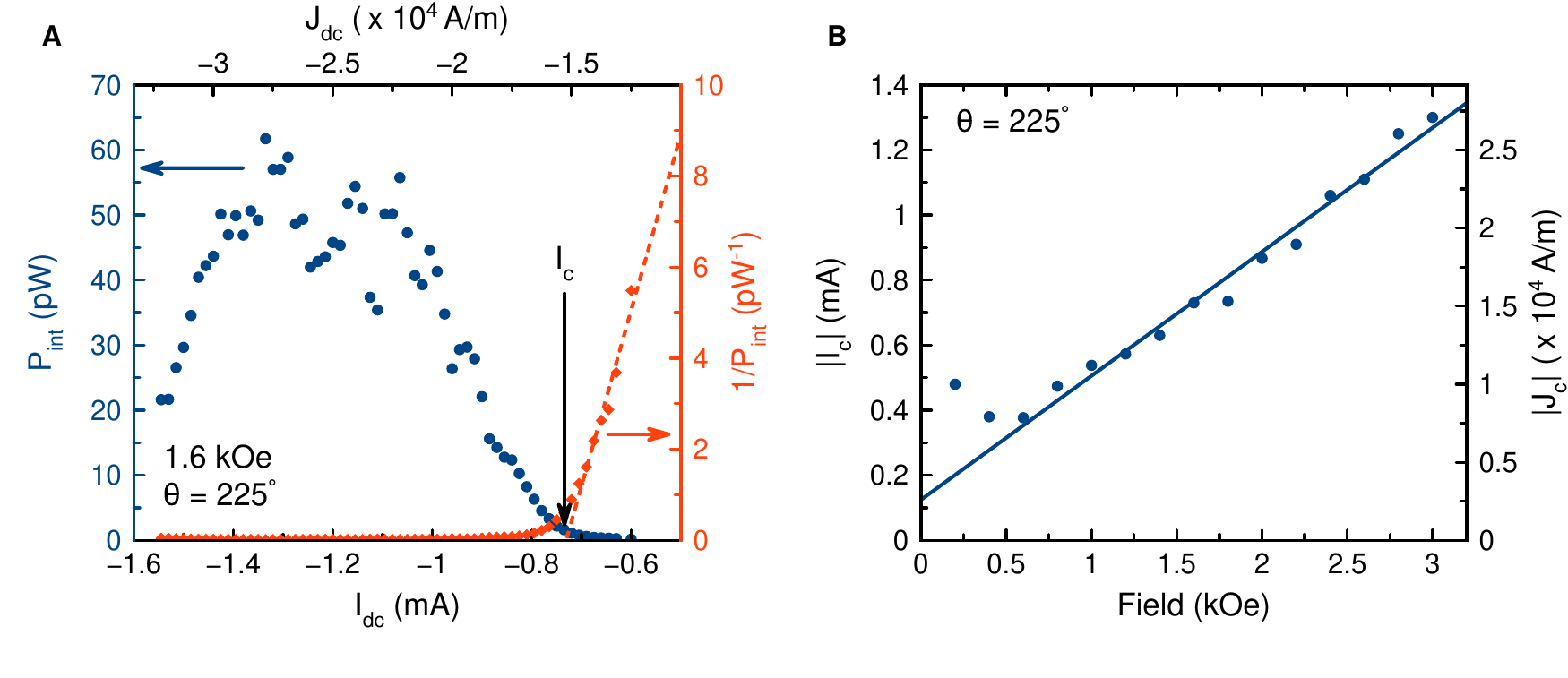}
\caption{
(\textbf{A})~Integrated power $P_\mathrm{int}$ emitted by the Ta(3\,nm)/Au(3.9\,nm)/FM/Ta(4\,nm) nanowire device in its lowest spin wave mode $M_1$ at $T=77$\,K (blue symbols) as a function of current bias measured in a 1.6\,kOe magnetic field applied at $\theta=225^\circ$. The inverse power $1/P_\mathrm{int}$ is shown by red symbols. (\textbf{B})~Absolute value of the critical current for the onset of the auto-oscillations $|I_\mathrm{c}|$  as a function of magnetic field applied at $\theta=225^\circ$.}
\label{emission_fits}
\end{figure*}

When direct current $I_\mathrm{dc}$ applied to the nanowire spin torque nano-oscillator (STNO) exceeds a critical value $I_\mathrm{c}$, magnetization of the FM layer enters an auto-oscillatory state, which gives rise to microwave power generation by the STNO device. We find that several spin wave eigenmodes of the nanowire enter the auto-oscillatory state, however most of the microwave power is generated by auto-oscillations of the lowest-frequency spin wave mode $M_1$. The integrated power $P_\mathrm{int}$ (blue symbols) emitted by mode $M_1$ in the Ta(3\,nm)/Au(3.9\,nm)/FM/Ta(4\,nm) nanowire device is shown in Fig.\,\ref{emission_fits}A. In this report, the power is expressed as power delivered to a $50\,\Omega$ load and is corrected for frequency dependent attenuation and amplification in the microwave measurement circuit. The data in Fig.\,\ref{emission_fits}A are collected at $T=77$\,K in a 1.6\,kOe magnetic field applied at $\theta=225^\circ$ in the $xz$ plane. We observe a large increase in the emitted microwave power above the background level for  $I_\mathrm{dc}<-0.7$\,mA. A more precise evaluation of $I_\mathrm{c}$ can be obtained by fitting the inverse of the integrated power (red symbols in Fig.\,\ref{emission_fits}A) to a straight line for sub-critical values of $I_\mathrm{dc}$ \cite{Slavin2009}. The critical current obtained via such a procedure from the data in Fig.\,\ref{emission_fits}A is $I_\mathrm{c}=-0.73$\,mA.

Figure\,\ref{emission_fits}B shows the dependence of $I_\mathrm{c}$ on magnetic field $H$ applied at $\theta=225^\circ$ in the $xz$ plane. Apart from the low field regime, this dependence is well fit by a straight line. This linear dependence of $I_\mathrm{c}$ on $H$ is expected for our devices because magnetic anisotropy of the nanowire is relatively small
\cite{Ralph2008,Slavin2009}. In the low field regime, magnetization of the nanowire starts to deviate from the applied field direction that maximizes the PHT efficiency ($\theta=225^\circ$), which leads to an increase of $I_\mathrm{c}$.

The microwave power emitted by this STNO in the $M_1$ mode is approximately 60\,pW at $I_{\mathrm{dc}}=-1.3$\,mA, which significantly exceeds the output power of spin Hall STNOs measured at higher currents \cite{Duan2014}. The reason for the improved efficiency of PHT STNO is the angular dependence of AMR, which maximizes the amplitude of resistance oscillations for magnetization making a $45^\circ$ angle with respect to the electric current direction. This magnetization direction in the $xz$ plane is also the direction of maximum efficiency of the PHT. This is to be contrasted with spin Hall STNOs, for which the direction of maximum negative damping efficiency ($\uvec{y}$) minimizes the amplitude of AMR oscillations.

\section{Relation between planar Hall torque and anisotropic magnetoresistance}

Both PHT and AMR arise from the planar Hall current in the FM layer, and thus PHT is expected to increase with increasing AMR. Since AMR in our NM/FM systems strongly depends on temperature, we can test the relation between PHT and AMR by measuring their temperature dependence. 

\begin{figure*}[!h]
\includegraphics[width=\textwidth]{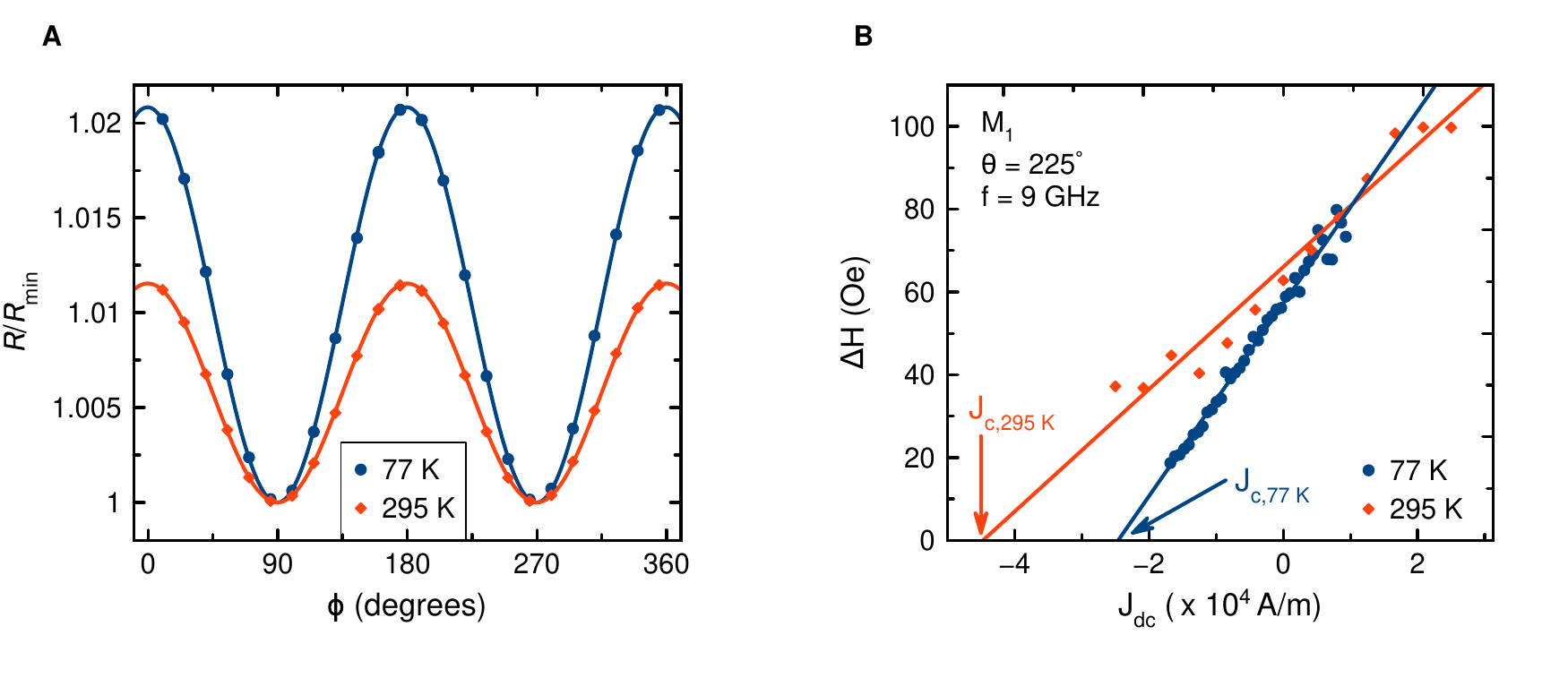}
\caption{
(\textbf{A})~Normalized resistance of the Ta(3\,nm)/Au(3.9\,nm)/FM/Ta(4\,nm) film as a function of 4\,kOe magnetic field angle applied in the $xy$ plane at 77\,K (blue symbols) and 295\,K (red symbols).(\textbf{B})~ST-FMR linewidth of the $M_1$ mode as a function of direct sheet current density $J_\mathrm{dc}$ applied to the Ta(3\,nm)/Au(3.9\,nm)/FM/Ta(4\,nm) nanowire device at $T=77$\,K (blue symbols) and $T=295$\,K (red symbols).   }
\label{temp_props}
\end{figure*}

The angular dependence of normalized resistance of a Ta(3\,nm)/Au(3.9\,nm)/FM/Ta(4\,nm) multilayer as a function of the direction of a 4\,kOe saturating magnetic field applied in the plane of the sample is shown in Fig.\,\ref{temp_props}A. The data reveal that the AMR ratio $\Delta \rho_{\rm{AMR}}/\rho$ of the multilayer increases by 80\% at $T=77$\,K compared to its room temperature value: $\left( \Delta \rho_{\rm{AMR}}/\rho \right)_{\rm{295\,K}} = 0.0116$; $\left( \Delta \rho_{\rm{AMR}}/\rho \right)_{\rm{77\,K}} = 0.0209$:
\begin{equation} 
\label{eq:AMRratio}
\frac{\left( \Delta \rho_{\rm{AMR}}/\rho \right)_{\rm{77\,K}}}{\left( \Delta \rho_{\rm{AMR}}/\rho \right)_{\rm{295\,K}}}=1.80.
\end{equation}

Fig.\,\ref{temp_props}B shows the critical sheet current density for the onset of auto-oscillations $J_{\rm{c}}$ measured by ST-FMR in the Ta(3\,nm)/Au(3.9\,nm)/FM/Ta(4\,nm) nanowire device at $T=295$\,K and $T=77$\,K. This $J_{\rm{c}}$ measured at 9\,GHz decreases from $J_{\rm{c,295\,K}}=-4.4 \times 10^{4}$\,A/m at room temperature to $J_{\rm{c,77\,K}}=-2.5 \times 10^{4}$\,A/m at 77\,K. The value of $J_{\rm{c,77\,K}}$ measured by ST-FMR (at 9\,GHz  corresponding to $H=2.83$\,kOe) is in agreement with the value of $J_{\rm{c}}$ obtained from the microwave emission measurements at this magnetic field value, as shown in Fig.\,\ref{emission_fits}B. The observed temperature-induced variation of $J_{\rm{c}}$ is consistent with the expected increase of the PHT efficiency due to the low-temperature increase of the planar Hall current and AMR.

The critical sheet current density can be approximated by \cite{Ralph2008}:
\begin{equation} 
\label{eq:Jc}
J_{\mathrm{c}} \approx \frac{2e}{\hbar} \frac{\alpha}{\eta} d M_s \frac{\omega}{\gamma}
\end{equation}
where $e$ is the electron charge, $\hbar$ is the reduced Planck's constant, $d$ is the FM layer thickness, $\omega$ is the oscillation angular frequency, $\gamma$ is the gyromagnetic ratio, and $\eta$ is a PHT efficiency parameter characterizing conversion of the sheet current density $J_\mathrm{dc}$ into the spin current density $Q^z_\mathrm{PHE}$.

The ratio of the PHT efficiencies at $T=77$\,K and $T=295$\,K can be determined from Eq.~(\ref{eq:Jc}):
\begin{equation} 
\label{eq:effratio}
\frac{\eta_\mathrm{77\,K}}{\eta_\mathrm{295\,K}} = \frac{J_\mathrm{c,295\,K}}{J_\mathrm{c,77\,K}} \frac{\alpha_\mathrm{77\,K}}{\alpha_\mathrm{295\,K}},
\end{equation}
where we neglect the temperature dependence of $ M_{\rm{s}} $ since it was shown to change by less than $4\%$ between 295\,K and 4\,K \cite{Kurt2010b,Zhang2014,Arora2017a}. We employ broadband ferromagnetic resonance to measure the Gilbert damping parameter $\alpha$ at $T=295$\,K and $T=77$\,K and find $\alpha_{\rm{295\,K}}=20.9 \times 10^{-3}$ while $\alpha_{\rm{77\,K}} = 22.2 \times 10^{-3}$. Substituting the measured values of $J_\mathrm{c}$ and $\alpha$ into Eq.~(\ref{eq:effratio}), we obtain:
\begin{equation} 
\label{eq:ERvalue}
\frac{\eta_\mathrm{77\,K}}{\eta_\mathrm{295\,K}} = 1.9 \pm 0.1.
\end{equation}
Comparison of Eq.~(\ref{eq:Jc}) and Eq.~(\ref{eq:ERvalue}) shows that the PHT efficiency scales with the AMR ratio, which supports the planar Hall current  origin of PHT.


\begin{thebibliography}{10}

\bibitem{Ando2008}
K.~Ando, S.~Takahashi, K.~Harii, K.~Sasage, J.~Ieda, S.~Maekawa, E.~Saitoh,
  {\it Phys. Rev. Lett.\/} {\bf 101}, 036601 (2008).

\bibitem{Miron2011}
I.~M. Miron, K.~Garello, G.~Gaudin, P.-J. Zermatten, M.~V. Costache,
  S.~Auffret, S.~Bandiera, B.~Rodmacq, A.~Schuhl, P.~Gambardella, {\it
  Nature\/} {\bf 476}, 189 (2011).

\bibitem{Liu2012}
L.~Liu, C.-F. Pai, Y.~Li, H.~W. Tseng, D.~C. Ralph, R.~A. Buhrman, {\it
  Science\/} {\bf 336}, 555 (2012).

\bibitem{Liu2012b}
L.~Liu, C.-F. Pai, D.~C. Ralph, R.~A. Buhrman, {\it Phys. Rev. Lett.\/} {\bf
  109}, 186602 (2012).

\bibitem{Demidov2012}
V.~E. Demidov, S.~Urazhdin, H.~Ulrichs, V.~Tiberkevich, A.~Slavin, D.~Baither,
  G.~Schmitz, S.~O. Demokritov, {\it Nat. Mater.\/} {\bf 11}, 1028 (2012).

\bibitem{Duan2014}
Z.~Duan, A.~Smith, L.~Yang, B.~Youngblood, J.~Lindner, V.~E. Demidov, S.~O.
  Demokritov, I.~N. Krivorotov, {\it Nat. Commun.\/} {\bf 5}, 5616 (2014).

\bibitem{Collet2016}
M.~Collet, X.~de~Milly, O.~d'Allivy Kelly, V.~V. Naletov, R.~Bernard,
  P.~Bortolotti, J.~{Ben Youssef}, V.~E. Demidov, S.~O. Demokritov, J.~L.
  Prieto, M.~Mu{\~{n}}oz, V.~Cros, A.~Anane, G.~de~Loubens, O.~Klein, {\it Nat.
  Commun.\/} {\bf 7}, 10377 (2016).

\bibitem{Awad2017}
A.~A. Awad, P.~D{\"{u}}rrenfeld, A.~Houshang, M.~Dvornik, E.~Iacocca, R.~K.
  Dumas, J.~{\AA}kerman, {\it Nat. Phys.\/} {\bf 13}, 292 (2016).

\bibitem{Slavin2009}
A.~Slavin, V.~Tiberkevich, {\it IEEE Trans. Magn.\/} {\bf 45}, 1875 (2009).

\bibitem{Sinova2015}
J.~Sinova, S.~O. Valenzuela, J.~Wunderlich, C.~H. Back, T.~Jungwirth, {\it Rev.
  Mod. Phys.\/} {\bf 87}, 1213 (2015).

\bibitem{Manchon2015}
A.~Manchon, H.~C. Koo, J.~Nitta, S.~M. Frolov, R.~A. Duine, {\it Nat. Mater.\/}
  {\bf 14}, 871 (2015).

\bibitem{Fan2014}
X.~Fan, H.~Celik, J.~Wu, C.~Ni, K.-j. Lee, V.~O. Lorenz, J.~Q. Xiao, {\it Nat.
  Commun.\/} {\bf 5}, 3042 (2014).

\bibitem{MacNeill2016a}
D.~MacNeill, G.~M. Stiehl, M.~H.~D. Guimaraes, R.~A. Buhrman, J.~Park, D.~C.
  Ralph, {\it Nat. Phys.\/} {\bf 13}, 300 (2016).

\bibitem{Kokado2012}
S.~Kokado, M.~Tsunoda, K.~Harigaya, A.~Sakuma, {\it J. Phys. Soc. Japan\/} {\bf
  81}, 024705 (2012).

\bibitem{methods}
{\it Materials and methods are available as supplementary materials\/} .

\bibitem{Mangin2006}
S.~Mangin, D.~Ravelosona, J.~A. Katine, M.~J. Carey, B.~D. Terris, E.~E.
  Fullerton, {\it Nat. Mater.\/} {\bf 5}, 210 (2006).

\bibitem{Arora2017a}
M.~Arora, R.~H{\"{u}}bner, D.~Suess, B.~Heinrich, E.~Girt, {\it Phys. Rev. B\/}
  {\bf 96}, 024401 (2017).

\bibitem{Goncalves2013}
A.~M. Gon{\c{c}}alves, I.~Barsukov, Y.-J. Chen, L.~Yang, J.~A. Katine, I.~N.
  Krivorotov, {\it Appl. Phys. Lett.\/} {\bf 103}, 172406 (2013).

\bibitem{Taniguchi2015}
T.~Taniguchi, J.~Grollier, M.~D. Stiles, {\it Phys. Rev. Appl.\/} {\bf 3},
  044001 (2015).

\bibitem{Garello2013}
K.~Garello, I.~M. Miron, C.~O. Avci, F.~Freimuth, Y.~Mokrousov,
  S.~Bl{\"{u}}gel, S.~Auffret, O.~Boulle, G.~Gaudin, P.~Gambardella, {\it Nat.
  Nanotechnol.\/} {\bf 8}, 587 (2013).

\bibitem{Mosendz2010}
O.~Mosendz, V.~Vlaminck, J.~E. Pearson, F.~Y. Fradin, G.~E.~W. Bauer, S.~D.
  Bader, A.~Hoffmann, {\it Phys. Rev. B\/} {\bf 82}, 214403 (2010).

\bibitem{Humphries2017}
A.~M. Humphries, T.~Wang, E.~R.~J. Edwards, S.~R. Allen, J.~M. Shaw, H.~T.
  Nembach, J.~Q. Xiao, T.~J. Silva, X.~Fan, {\it Nat. Commun.\/} {\bf 8}, 911
  (2017).

\bibitem{Amin2016b}
V.~P. Amin, M.~D. Stiles, {\it Phys. Rev. B\/} {\bf 94}, 104420 (2016).

\bibitem{Saidaoui2016}
H.~B.~M. Saidaoui, A.~Manchon, {\it Phys. Rev. Lett.\/} {\bf 117}, 036601
  (2016).

\bibitem{Bender2016}
S.~A. Bender, Y.~Tserkovnyak, {\it Phys. Rev. B\/} {\bf 93}, 064418 (2016).

\bibitem{Safranski2016b}
C.~Safranski, I.~Barsukov, H.~K. Lee, T.~Schneider, A.~A. Jara, A.~Smith,
  H.~Chang, K.~Lenz, J.~Lindner, Y.~Tserkovnyak, M.~Wu, I.~N. Krivorotov, {\it
  Nat. Commun.\/} {\bf 8}, 117 (2017).

\bibitem{Mann2017}
M.~Mann, G.~S.~D. Beach, {\it APL Mater.\/} {\bf 5}, 106104 (2017).

\bibitem{Ryu2013}
K.-S. Ryu, L.~Thomas, S.-H. Yang, S.~Parkin, {\it Nat. Nanotechnol.\/} {\bf 8},
  527 (2013).

\bibitem{Duttagupta2017a}
S.~DuttaGupta, T.~Kanemura, C.~Zhang, A.~Kurenkov, S.~Fukami, H.~Ohno, {\it
  Appl. Phys. Lett.\/} {\bf 111}, 182412 (2017).

\bibitem{Legrand2017}
W.~Legrand, D.~Maccariello, N.~Reyren, K.~Garcia, C.~Moutafis,
  C.~Moreau-Luchaire, S.~Collin, K.~Bouzehouane, V.~Cros, A.~Fert, {\it Nano
  Lett.\/} {\bf 17}, 2703 (2017).

\bibitem{Arora2017}
M.~Arora, N.~R. Lee-Hone, T.~Mckinnon, C.~Coutts, R.~H{\"{u}}bner, B.~Heinrich,
  D.~M. Broun, E.~Girt, {\it J. Phys. D. Appl. Phys.\/} {\bf 50}, 505003
  (2017).

\bibitem{Mcguire1975}
T.~McGuire, R.~Potter, {\it IEEE Trans. Magn.\/} {\bf 11}, 1018 (1975).

\bibitem{Daalderop1992}
G.~H.~O. Daalderop, P.~J. Kelly, F.~J.~A. den Broeder, {\it Phys. Rev. Lett.\/}
  {\bf 68}, 682 (1992).

\bibitem{Miccoli2015}
I.~Miccoli, F.~Edler, H.~Pfn{\"{u}}r, C.~Tegenkamp, {\it J. Phys. Condens.
  Matter\/} {\bf 27}, 223201 (2015).

\bibitem{Thorsteinsson2009}
S.~Thorsteinsson, F.~Wang, D.~H. Petersen, T.~M. Hansen, D.~Kj{\ae}r, R.~Lin,
  J.-Y. Kim, P.~F. Nielsen, O.~Hansen, {\it Rev. Sci. Instrum.\/} {\bf 80},
  053902 (2009).

\bibitem{Montoya2016}
E.~Montoya, P.~Omelchenko, C.~Coutts, N.~R. Lee-Hone, R.~H{\"{u}}bner,
  D.~Broun, B.~Heinrich, E.~Girt, {\it Phys. Rev. B\/} {\bf 94}, 054416 (2016).

\bibitem{Montoya2014}
E.~Montoya, T.~McKinnon, A.~Zamani, E.~Girt, B.~Heinrich, {\it J. Magn. Magn.
  Mater.\/} {\bf 356}, 12 (2014).

\bibitem{Montoya2016a}
E.~Montoya, T.~Sebastian, H.~Schultheiss, B.~Heinrich, R.~E. Camley,
  Z.~Celinski, {\it Magn. Surfaces, Interfaces, Nanoscale Mater.\/}, R.~E.
  Camley, Z.~Celinski, R.~L. Stamps, eds. (Elsevier B.V., 2015), vol.~5,
  chap.~3, pp. 113--167, first edn.

\bibitem{Heinrich1993}
B.~Heinrich, J.~Cochran, {\it Adv. Phys.\/} {\bf 42}, 523 (1993).

\bibitem{McMichael2003}
R.~D. McMichael, D.~J. Twisselmann, A.~Kunz, {\it Phys. Rev. Lett.\/} {\bf 90},
  227601 (2003).

\bibitem{Bland2005}
B.~Heinrich, {\it Ultrathin Magn. Struct.\/}, J.~A.~C. Bland, B.~Heinrich, eds.
  (Springer-Verlag, Berlin/Heidelberg, 2005), vol. III, chap.~5, pp. 143--206.

\bibitem{Urban2001}
R.~Urban, G.~Woltersdorf, B.~Heinrich, {\it Phys. Rev. Lett.\/} {\bf 87},
  217204 (2001).

\bibitem{Simanek2003}
E.~{\v{S}}im{\'{a}}nek, B.~Heinrich, {\it Phys. Rev. B\/} {\bf 67}, 144418
  (2003).

\bibitem{Tserkovnyak2005}
Y.~Tserkovnyak, A.~Brataas, G.~E.~W. Bauer, B.~I. Halperin, {\it Rev. Mod.
  Phys.\/} {\bf 77}, 1375 (2005).

\bibitem{Hahn2013}
C.~Hahn, G.~de~Loubens, O.~Klein, M.~Viret, V.~V. Naletov, J.~{Ben Youssef},
  {\it Phys. Rev. B\/} {\bf 87}, 174417 (2013).

\bibitem{Mizukami2001}
S.~Mizukami, Y.~Ando, T.~Miyazaki, {\it J. Magn. Magn. Mater.\/} {\bf 226-230},
  1640 (2001).

\bibitem{Singh2017a}
B.~B. Singh, S.~K. Jena, S.~Bedanta, {\it J. Phys. D. Appl. Phys.\/} {\bf 50},
  345001 (2017).

\bibitem{Tulapurkar2005}
A.~A. Tulapurkar, Y.~Suzuki, A.~Fukushima, H.~Kubota, H.~Maehara, K.~Tsunekawa,
  D.~D. Djayaprawira, N.~Watanabe, S.~Yuasa, {\it Nature\/} {\bf 438}, 339
  (2005).

\bibitem{Sankey2006}
J.~C. Sankey, P.~M. Braganca, A.~G.~F. Garcia, I.~N. Krivorotov, R.~A. Buhrman,
  D.~C. Ralph, {\it Phys. Rev. Lett.\/} {\bf 96}, 227601 (2006).

\bibitem{Chiba2015}
T.~Chiba, M.~Schreier, G.~E.~W. Bauer, S.~Takahashi, {\it J. Appl. Phys.\/}
  {\bf 117}, 17C715 (2015).

\bibitem{Ando2009}
K.~Ando, J.~Ieda, K.~Sasage, S.~Takahashi, S.~Maekawa, E.~Saitoh, {\it Appl.
  Phys. Lett.\/} {\bf 94}, 262505 (2009).

\bibitem{Ralph2008}
D.~Ralph, M.~Stiles, {\it J. Magn. Magn. Mater.\/} {\bf 320}, 1190 (2008).

\bibitem{Kurt2010b}
H.~Kurt, M.~Venkatesan, J.~M.~D. Coey, {\it J. Appl. Phys.\/} {\bf 108}, 073916
  (2010).

\bibitem{Zhang2014}
P.~Zhang, K.~Xie, W.~Lin, D.~Wu, H.~Sang, {\it Appl. Phys. Lett.\/} {\bf 104},
  082404 (2014).

\end{thebibliography}
\end{document}